\documentclass [11pt]{article}
\title{Analysis of the Anomalous Brillet and Hall\\Experimental Result}
\author{Robert D. Klauber\\1100 University Manor Dr., 38B, Fairfield, Iowa 52556\\klauber@.iowatelecom.net, permanent: rklauber@.netscape.net\\  \\Published: \textit{Found. Phys. Lett. } June 2004}
\date{  }

\usepackage{graphicx}
\usepackage{longtable}

\begin{document}

\maketitle

\begin{abstract}

The persistent, second order, anomalous signal found in the Brillet and Hall 
experiment is derived by applying 4D differential geometry in the rotating 
earth frame. By incorporating the off diagonal time-space components of the 
rotating frame metric directly into the analysis, rather than arbitrarily 
transforming them away, one finds a signal dependence on the surface speed 
of the earth due to rotation about its axis. This leads to a Brillet-Hall 
signal prediction in remarkably close agreement with experiment. No signal 
is predicted from the speed of the earth in solar or galactic orbit, as the 
associated metric for gravitational orbit has no off diagonal component. To 
corroborate this result, a repetition by other experimentalists of the 
Brillet-Hall experiment, in which the test apparatus turns with respect to 
the earth surface, is urged.

\end{abstract}

\smallskip


\bigskip

\section{INTRODUCTION}
\label{sec:introduction}
\subsection{Background}
In 1978, Brillet and Hall\cite{Brillet:1979} used a 
Fabry-Perot\cite{Useful:1} interferometer that rotated with respect to the 
lab to measure the isotropy of space (i.e., of the speed of light). As the 
apparatus turned, anisotropic speed of light would cause otherwise coherent 
light waves to change phase and partially interfere, thereby reducing the 
transmitted wave intensity. A servo adjusted the frequency (and therefore 
the wavelength and phase) to maximize the transmitted wave intensity. Thus, 
the change in frequency was a direct measure of the degree of anisotropy.

Brillet-Hall found a "null" effect at the $\Delta f / f $= 3X10$^{-15}$ level, 
where $f$ is the frequency of the electromagnetic wave employed and $\Delta 
f$ is the amplitude of the change in $f$ as the apparatus is rotated. However, 
to obtain this result they subtracted out an anomalous non-null signal of 
approximate amplitude 2X10$^{-13}$ at twice the apparatus rotation rate. 
This signal had effectively constant phase in the laboratory (earth) frame, 
and though ``persistent'', was deemed ``spurious'' without further comment.

In 1981, Aspden\cite{Aspden:1981} noted that this anomalous signal 
corresponded closely to what pre-relativity physicists would have expected 
solely from the earth surface velocity. According to logic employed by 
Michelson and Morley\cite{See:1965}, $\Delta f / f$ is related to 
$\raise.5ex\hbox{$\scriptstyle 1$}\kern-.1em/ 
\kern-.15em\lower.25ex\hbox{$\scriptstyle 2$} v^{2}/c^{2}$, where $v$ is the 
speed of the test apparatus relative to the frame in which light speed is 
isotropic. Aspden suggested that perhaps in some little understood way, the 
earth surface speed might be somehow different in nature than the solar and 
galactic orbit speeds. His analysis implied that, if light speed were 
anisotropic, $\Delta f$/$f$ would have an amplitude of 
(.262)($\raise.5ex\hbox{$\scriptstyle 1$}\kern-.1em/ 
\kern-.15em\lower.25ex\hbox{$\scriptstyle 2$} v^{2}/c^{2})$ . Setting this 
result equal to the measured signal of $\sim $2X10$^{-13}$, one finds a $v$ 
value of approximately 362 m/sec. The earth surface velocity at the test 
site due to rotation about the earth's axis is 355 m/sec.

The present author believes Aspden's analysis to be flawed\footnote{ Aspden 
considered the change in angle of the output signal due to ``motion of the 
mirror relative to the light reference frame'' to be responsible for the 
non-null signal variation observed. However, the output signal in the 
Brillet-Hall experiment was monitored only for frequency change, and for an 
observer and apparatus fixed in the same frame, this does not vary with 
frame orientation, even in a Galilean analysis. (Wave speed and wavelength 
would change, but not frequency.) See Section \ref{subsec:fabry} 
herein, ref. \cite{Aspden:1981}, and ref. 
\cite{Klauber:1}.}\cite{Klauber:1}. Nevertheless, the detected signal 
is within a factor of two of what Michelson-Morley would have attributed to 
the earth surface velocity. (It turns out, as we will see, that 
$\raise.5ex\hbox{$\scriptstyle 1$}\kern-.1em/ 
\kern-.15em\lower.25ex\hbox{$\scriptstyle 2$} v^{2}/c^{2}$ would be 
peak-to-peak in the Brillet-Hall test, and half of this would be the 
amplitude.) In the present article, this signal is evaluated using general 
relativity/differential geometry in which off diagonal time-space components 
occur in the rotating (earth) frame metric. For such a metric, time is not 
orthogonal to space, and this is referred to as non-time-orthogonality. As 
shown below, non-time-orthogonal (NTO) 
analysis\cite{Klauber:1998}$^{,}$\cite{Klauber:2} implies an anisotropy in 
the speed of light in rotating frames, which when applied to the earth, 
leads to a prediction of the observed $\sim $2X10$^{-13}$ signal.

In this context, we note that as a result of studies on global positioning 
system (GPS) electromagnetic signal data, recognized world leading GPS 
expert Neil Ashby recently noted in \textit{Physics Today},

\textit{`` .. the principle of the constancy of c cannot be applied in a rotating reference frame ..'' }\cite{Ashby:2002}. 

He has also stated,

\textit{``Now consider a process in which observers in the rotating frame attempt to use Einstein synchronization }[constancy of the speed of light]\textit{ ..... Simple minded use of Einstein synchronization in the rotating frame ... thus leads to a significant error''.}\cite{Ashby:1997}

It is emphasized preemptively that NTO analysis does not contravene the 
geometric foundation of relativity theory. In fact, it is a direct 
consequence of straightforward application of differential geometry to cases 
where time is not orthogonal to space. These include rotating frames and 
spacetime around a star possessing angular momentum (in which off diagonal 
space-time metric components also exist.) NTO methodology leaves unchanged 
all analyses of systems in which time is orthogonal to space. Such systems 
comprise free fall frames in gravitational orbits and the vast majority of 
all other relativistic systems. Hence, the null signals obtained in many 
tests for effects of the solar or galactic orbit speeds are not in conflict 
with NTO analysis.

Further, NTO analysis is in complete agreement with all known 
experiments\cite{Klauber:3}$^{,}$\cite{Klauber:4}$^{,}$\cite{Klauber:5}$^{,}$\cite{Klauber:2003}. 
Of these experiments, only the Brillet-Hall test has been capable of 
monitoring any effect due to non-time-orthogonality of the rotating earth 
fixed reference frame\footnote{ Arguments can, however, be made that the 
Sagnac effect (see ref. \cite{Klauber:2003}) results from such 
non-time-orthogonality.}.

\subsection{Analysis Approach}
\label{subsec:analysis}
The steps in analysis presented herein are as follows.

\begin{enumerate}
\item Introductory overview of NTO analysis (Section \ref{sec:overview}.)
\item Determination of physical light speed (i.e., measured with physical instruments) in the circumferential and radial directions of a rotating frame via NTO analysis. (Section \ref{subsec:light}), and from them, the difference in time delay for the two directions (Sections \ref{subsec:round}, \ref{subsec:mylabel2}, and \ref{subsec:rotation}.)
\item A review of the Brillet and Hall apparatus, analysis of the result predicted by the NTO approach due to differences in time delay of light signals in the radial and circumferential direction, and a comparison with the actual results. (Section \ref{sec:brillet}.) The signal predicted is seen to be of the same order of magnitude but off by somewhat less than a factor of two.
\item An analysis of effects on the light signal other than time delay predicted by the NTO approach. (Section \ref{sec:further}.) These other effects are transverse to the light propagation direction, whereas the signal time delay is a parallel direction effect. The two combined are then shown to result in a predicted signal strikingly close to that measured by Brillet and Hall.
\item An explanation for why null signals are predicted by NTO analysis for solar and galactic orbital speeds but not for the earth surface speed due to rotation about its axis. (Section \ref{sec:gravitational}.)
\item Summary and Conclusions (Section \ref{sec:summary}.)
\end{enumerate}

\section{OVERVIEW OF NTO ANALYSIS }
\label{sec:overview}
The most widely accepted transformation\footnote{ There is significant 
background behind this transformation that is discussed in greater depth in 
refs \cite{Klauber:1998} and \cite{Klauber:2}. We shall tentatively 
accept this transformation as valid and find in Section 
\ref{sec:further} that it results in a correct prediction for the 
experiment under consideration.} from the non-rotating (lab, upper case 
symbols) frame to a rotating (lower case) frame is
\begin{equation}
\label{eq1}
\begin{array}{l}
 cT=ct \\ 
 R=r \\ 
 \Phi =\phi +\omega t \\ 
 Z=z \\ 
 \end{array}
\end{equation}
where $\omega $ is the angular velocity of the rotating frame and 
cylindrical spatial coordinates are used. The coordinate time $t$ for the 
rotating system equals the proper time of a standard clock located in the 
lab, which equals the time on a standard clock at the origin (axis of 
rotation) of the rotating frame.

Substituting the differential form of (\ref{eq1}) into the line element in the lab 
frame
\begin{equation}
\label{eq2}
ds^2\mbox{ }=\mbox{ }-\mbox{ }c^2dT^2+\mbox{ }dR^2\mbox{ }+\mbox{ }R^2d\Phi 
^2\mbox{ }+\mbox{ }dZ^2
\end{equation}
results in the line element for the rotating frame\footnote{ The first 
person to determine this metric for rotating frames may have been Paul 
Langevin [14], who used it successfully to analyze the Sagnac experiment, 
though he did not suggest it would lead to a non-null Michelson-Morley 
result for rotation. See further comments in this regard by the present 
author in refs [6], [7], and [13].}\cite{Langevin:1937}
\begin{equation}
\label{eq3}
ds^2\;\;=\;-c^2(1-\textstyle{{r^2\omega ^2} \over 
{c^2}})dt^2\;+\;dr^2\;+\;r^2d\phi ^2\;+\;2r^2\omega d\phi 
dt\;+\;dz^2=\;g_{\alpha \beta } dx^\alpha dx^\beta .
\end{equation}
Note that the metric in (\ref{eq3}) is not diagonal, since $g_{\phi t} \ne 0$, and 
this implies that time is not orthogonal to space (i.e., an NTO frame.) This 
mathematical feature becomes physically significant due to the fact that in 
a rotating frame it is not possible to synchronize clocks continuously 
without producing an NTO coordinatization. (See refs. [6] and [7].)

Time on a standard clock at any fixed 3D location in the rotating frame, 
found by taking \textit{ds}$^{2}= -c^{2}d\tau $ and \textit{dr = d}$\phi $ \textit{= dz = }0, is
\begin{equation}
\label{eq4}
d\tau =d\hat {t}=\sqrt {-g_{tt} } dt=\sqrt {1-r^2\omega ^2/c^2} dt,
\end{equation}
where the caret over \textit{dt} indicates \textit{physical} (proper) time (i.e., time measured with 
physical world standard clocks fixed in the rotating frame.) Obviously, time 
becomes imaginary for $r>c/\omega $, speed becomes faster than light, and 
physical clocks (or any material body) cannot exist at such locations.

\section{LIGHT SPEEDS AND TRAVEL TIMES}
\label{sec:light}
\subsection{Light Speeds in Two Directions}
\label{subsec:light}
\subsubsection{Circumferential Direction Light Speed}
For light \textit{ds}$^{2}$ = 0. Inserting this into (\ref{eq3}), taking \textit{dr=dz=}0, and using the 
quadratic equation formula, one obtains a light coordinate velocity 
(generalized coordinate spatial grid units per coordinate time unit) in the 
circumferential direction
\begin{equation}
\label{eq5}
v_{light,coord,circum} =\frac{d\phi }{dt}=-\omega \pm \frac{c}{r},
\end{equation}
where the sign before the last term depends on the circumferential direction 
of travel of the light ray. The physical velocity (the value one would 
measure in experiment using standard meter sticks and clocks in units of 
meters per second) is found from this to be$^{ }$\footnote{ The texts and 
articles listed in ref. \cite{Klauber:1972} are among those that discuss 
physical vector and tensor components (the values one measures in 
experiment) and the relationship between them and coordinate components (the 
mathematical values that depend on the generalized coordinate system being 
used.) This relationship is $u^{\hat {\mu }\,}\,=\sqrt {g_{\underline{\mu 
}\underline{\mu }} } u^\mu $ where the caret over the index indicates a 
physical vector component and underlining implies no 
summation.}\cite{Klauber:1972} 
\begin{equation}
\label{eq6}
v_{light,phys,circum} \;\;=\;\;\frac{\sqrt {g_{\phi \phi } } d\phi }{\sqrt 
{-g_{tt} } dt}\;\;=\;\;\frac{-\;r\omega \;\pm \;c}{\sqrt 
{1-\textstyle{{\omega ^2r^2} \over {c^2}}} }=\;\;\frac{-\;v\;\pm \;c}{\sqrt 
{1-\textstyle{{v^2} \over {c^2}}} },
\end{equation}
where $v=\omega r$ is the circumferential speed of a point fixed in the 
rotating frame as seen from the lab. Note that for rotation, the physical 
speed of light is not invariant or isotropic, and that this lack of 
invariance/isotropy depends on $\omega $, the angular velocity. If $\omega 
$=0, light speed is isotropic and invariant, the metric (\ref{eq3}) is diagonal, and 
time is orthogonal to space.

\subsubsection{Radial Direction Light Speed}
\label{subsubsec:radial}
For a radially directed ray of light, \textit{d$\phi $} = \textit{dz = }0, and \textit{ds} = 0. Solving (\ref{eq3}) for 
\textit{dr/dt} one obtains
\begin{equation}
\label{eq7}
\frac{dr}{dt}=c\sqrt {1-r^2\omega ^2/c^2} .
\end{equation}
Since $g_{rr}$ = 1, the physical component (measured with standard meter 
sticks) for radial displacement $d\hat {r}$ equals the coordinate radial 
displacement \textit{dr}. The physical (measured) speed of light in the radial 
direction is therefore
\begin{equation}
\label{eq8}
v_{light,phys,radial} =\frac{d\hat {r}}{d\hat {t}}=\frac{\sqrt {g_{rr} } 
dr}{\sqrt {-g_{tt} } dt}=\frac{dr}{\sqrt {1-r^2\omega ^2/c^2} dt}=c.
\end{equation}
For details on NTO analysis, the reader is referred to 
Klauber\cite{Klauber:1998}$^{,}$\cite{Klauber:2}.

\subsection{Round Trip Travel Times}
\label{subsec:round}
\subsubsection{Circumferential Round Trip Travel Time}
In a given direction the outward and return round trip time over a path 
length $l$ is 
\begin{equation}
\label{eq9}
t_{RT} =\frac{l}{v_+ }+\frac{l}{v_- },
\end{equation}
where notation should be obvious. Inserting (\ref{eq6}) into (\ref{eq9}), and using absolute 
values for speed in (\ref{eq6}), one obtains the circumferential round trip light 
time
\begin{equation}
\label{eq10}
t_{RT,circum} =\frac{l\sqrt {1-v^2/c^2} }{c-v}+\frac{l\sqrt {1-v^2/c^2} 
}{c+v}.
\end{equation}
This readily becomes
\begin{equation}
\label{eq11}
t_{RT,circum} =\frac{2lc\sqrt {1-v^2/c^2} }{c^2-v^2}=\frac{2l}{c\sqrt 
{1-v^2/c^2} }\cong \frac{2l}{c}\left( {1+\textstyle{1 \over 
2}\frac{v^2}{c^2}} \right).
\end{equation}
\subsubsection{Radial Round Trip Travel Time}
\label{subsubsec:mylabel2}
From (\ref{eq8}), we know that in rotating frames for the radial direction $v_{+}$ = 
$v_{-}=c$. Using these values in (\ref{eq9}), one finds the round trip (assuming 
$g_{tt}$ is effectively constant over short distances) time, as measured on 
standard clocks fixed in the rotating frame, to be
\begin{equation}
\label{eq12}
t_{RT,radial} =\frac{2l}{c}.
\end{equation}
This is no different from that of inertial systems in special relativity.

\subsection{Round Trip Time Difference}
\label{subsec:mylabel2}
Consider the earth as a rotating frame with either a Michelson-Morley or 
Brillet-Hall experiment mounted on the earth's surface. Each apparatus turns 
so that the perpendicular and parallel (to the earth surface speed around 
its axis) directions are alternately tested for the round trip light travel 
times.

The difference between the parallel and perpendicular directions for 
rotation is (\ref{eq11}) minus (\ref{eq12}),
\begin{equation}
\label{eq13}
\Delta t_{RT,NTO\;frame\;rotation} \cong \frac{2l}{c}\left( {\textstyle{1 
\over 2}\frac{v^2}{c^2}} \right)=(t_{RT,no\;motion} )\left( {\textstyle{1 
\over 2}\frac{v^2}{c^2}} \right)=\textstyle{1 \over 2}\beta 
^2(t_{RT,no\;motion} ).
\end{equation}
\subsection{Rotation vs Translation}
\label{subsec:rotation}
For the NTO rotating frame analysis, the radial direction corresponds to the 
perpendicular direction for a translating system. Note, the travel times for 
this direction between Galilean translation (reviewed in Appendix B) and NTO relativistic rotation 
are not the same, and differ by a factor of $\raise.5ex\hbox{$\scriptstyle 
1$}\kern-.1em/ \kern-.15em\lower.25ex\hbox{$\scriptstyle 2$} $ before the 
$\beta ^{2}$. (Compare (\ref{eq12}) to (\ref{eq46}).) 

Note also that the travel times between Galilean translation and NTO 
relativistic rotation in the parallel to velocity direction also differ by a 
factor of $\raise.5ex\hbox{$\scriptstyle 1$}\kern-.1em/ 
\kern-.15em\lower.25ex\hbox{$\scriptstyle 2$} $ before $\beta ^{2}$ 
(compare (\ref{eq11}) to (\ref{eq47}).) 

However, the difference between the parallel and perpendicular directions 
for both cases is (\ref{eq13}), i.e., the same as (\ref{eq48}). One can thus conclude that 
according to NTO analysis, the Michelson-Morley and Brillet-Hall experiments 
performed on a rotating frame (i.e., the earth) would detect signals of the 
same magnitude as predicted for Galilean translation.

\section{BRILLET-HALL}
\label{sec:brillet}
\subsection{Brillet-Hall's Fabry-Perot Interferometer}
\label{subsec:brillet}
A Fabry-Perot interferometer\cite{Useful:1} consists of two partially 
transmitting, partially reflecting mirrors separated by an ``etalon'' (of 
length $d)$ and facing one another as shown in Figure 1. Light entering the 
left side of the Fabry-Perot interferometer in Figure 1 emerges as a series 
of rays on the right side, and creates a fringe pattern. For reference, a 
derivation of the well known constructive interference relation (shown 
inside the box of Figure 1 and repeated below as (\ref{eq14})) is provided in 
Appendix A.

To help make a point later, we emphasize here that a wave front entering 
from the left is visualized as a collection of Huygen's type wavelets. The 
particular ray components of such wavelets entering the left glass section 
at angle $\alpha $ produce constructive interference at an exit angle of 
$\alpha $. Other ray components of the same wave front, entering at other 
angles, produce various degrees of destructive interference. In essence 
then, a change in angle of incidence of the centerline of a macroscopic beam 
of light would have no effect on the angle $\alpha $ at which constructive 
interference occurs.

\begin{figure}[htbp]
\centerline{\includegraphics[width=5.48in,height=3.15in]{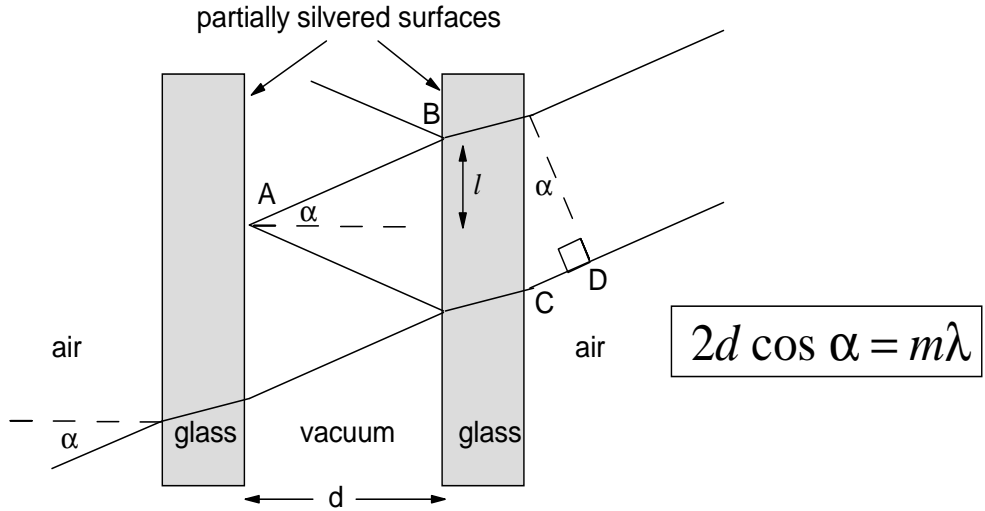}}
\caption{Fabry-Perot Interferometer}
\label{fig1}
\end{figure}

The angle $\alpha $ of peak intensity for light exiting the Fabry-Perot 
interferometer is found from
\begin{equation}
\label{eq14}
\cos \alpha =\frac{m\lambda }{2d},
\end{equation}
where $m$, the mode number, is an integer, and different values for $m$ correspond 
to different angles of local peaks in intensity.

\subsection{Fabry-Perot and the Brillet-Hall Experiment}
\label{subsec:fabry}
The incoming light ray depicted in the Fabry-Perot interferometer of Figure 
2 has an incident angle of zero. The upper part of the figure shows the 
right traveling incident ray (ray {\#}1), its left traveling reflection off 
of the RHS (ray {\#}1r), and the right traveling reflection of the ray 
{\#}1r (ray {\#}2). For simplicity, additional successive rays ({\#}2r, 
{\#}3, {\#}3r, etc.) that would result from reflection of ray {\#}2 and 
subsequent reflections are not shown. Isotropy of light speed is assumed in 
the upper part of Figure 2, and the wavelength has been adjusted (by varying 
the incident light frequency) so that ray {\#}1 and ray {\#}2 are in phase 
when they strike the RHS. The portions of {\#}1 and {\#}2 that exit the RHS 
will then combine to maximize intensity of the resultant beam. Higher order 
rays ({\#}3, {\#}4, etc. not shown) would then be in phase with {\#}1 and 
{\#}2 as well.

From (\ref{eq14}), the condition for constructive interference with $\alpha =0$ and 
isotropic light speed is
\begin{equation}
\label{eq15}
\lambda =\frac{2d}{m}.
\end{equation}
Note that a change in either the etalon length or the wavelength will reduce 
the constructive interference at $\alpha $ = 0 and thus, the intensity 
sensed.

\begin{figure}[htbp]
\centerline{\includegraphics[width=5.39in,height=3.46in]{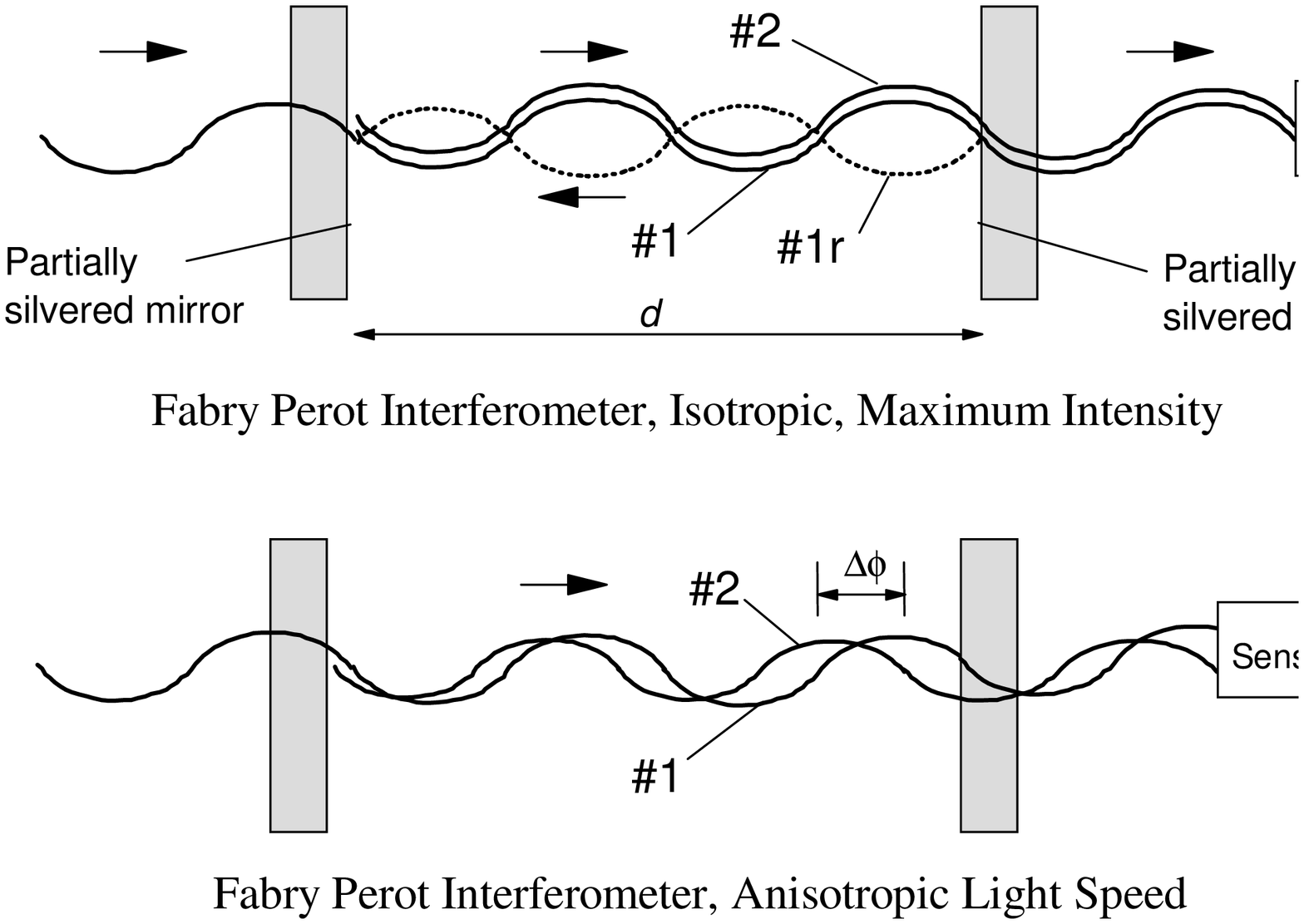}}
\caption{Anisotropic vs Isotropic Light Speed in the Fabry-Perot Interferometer}
\label{fig2}
\end{figure}

For anisotropic light speed (lower part of Figure 2), the round trip time 
for a given wave maximum to reflect from the right mirror and return to the 
right mirror would be greater than for the isotropic case. Hence, the phase 
relation between waves {\#}1 and {\#}2 would vary as the apparatus turned, 
and thus so would intensity of the light entering the sensor. By judiciously 
changing the wavelength as the orientation changed, one could, in principle, 
maintain maximum constructive interference.

Brillet-Hall used this property of the Fabry-Perot interferometer to measure 
light speed anisotropy. They employed a He Ne generated laser that was 
successively reflected off two mirrors and into an Fabry-Perot 
interferometer, as shown in Figure 3. A sensor at fixed angle relative to 
the Fabry-Perot interferometer monitored the output radiation intensity, and 
the signal from it was fed into a servo unit. The entire apparatus shown in 
Figure 3 was mounted on a 95 cm by 40 cm by 12 cm granite table. The table 
was then turned about an axis perpendicular to the table surface at the rate 
of one revolution every 10 seconds. As the apparatus turned, the servo unit 
continually adjusted the frequency (and hence the wavelength) of the 
generated laser to maintain sensor intensity at a maximum. Variation in this 
frequency was therefore a direct measure of the anisotropy of light speed.

\begin{figure}[htbp]
\centerline{\includegraphics[width=3.30in,height=3.29in]{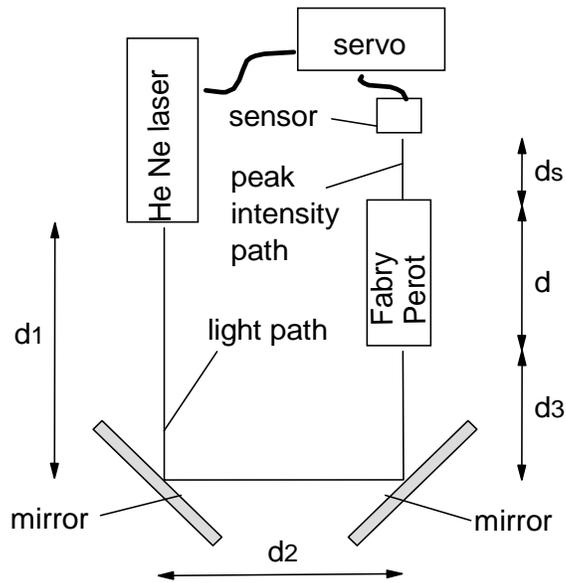}}
\caption{Schematic of the Brillet-Hall Apparatus}
\label{fig3}
\end{figure}

Thus, if light speed were anisotropic in rotating frames as NTO analysis 
suggests, the $\raise.5ex\hbox{$\scriptstyle 1$}\kern-.1em/ 
\kern-.15em\lower.25ex\hbox{$\scriptstyle 2$} \beta ^{2}$ of (\ref{eq13}) would 
represent the fractional change in time between maxima of the first and 
second right traveling waves as they exit the etalon on the RHS. That is, it 
represents the phase lag of the second wave behind the first upon arrival at 
the RHS. Changing the wavelength (done by changing the frequency) can reduce 
this phase lag to zero and maximize the detected intensity. Hence, the 
amount the servo changes frequency in order to maximize intensity indicates 
the magnitude of $\raise.5ex\hbox{$\scriptstyle 1$}\kern-.1em/ 
\kern-.15em\lower.25ex\hbox{$\scriptstyle 2$} \beta ^{2}$ and therefore 
the magnitude of $v$ detectable (i.e., the degree of anisotropy.)

\subsection{Time Delay Due to Anisotropy in the Brillet-Hall Test}
\label{subsec:mylabel1}
We saw in Section \ref{subsec:rotation} that the time delay 
variation imparted to wave {\#}2 in Figure 2 as a result of changing 
orientation of the Fabry-Perot interferometer through 90$^{o}$ for both 
translating frames in a Galilean universe and NTO rotating frames is (\ref{eq13}), 
i.e.,
\begin{equation}
\label{eq16}
\Delta t_{RT,moving\,frame} \cong \textstyle{1 \over 2}\beta 
^2(t_{RT,no\;motion} ).
\end{equation}
Note, however, that the Michelson-Morley apparatus has two perpendicular 
legs and results in (\ref{eq16}) representing a single-sided amplitude variation in 
the fringing as the apparatus turns. This is a result of the lead time of 
one leg over the other becoming a lag time when the apparatus is turned 
90$^{o}$. In the Brillet-Hall experiment, on the other hand, the round trip 
time difference, although numerically equal to (\ref{eq16}), is a peak-to-peak 
(double-sided) amplitude. The Brillet-Hall apparatus uses a Fabry-Perot 
interferometer as a single arm and the time delay varies between zero for 
the radial direction and the same value (given by (\ref{eq11})) for both the 
positive and negative circumferential directions.

\subsection{Brillet-Hall: NTO Time Delay Prediction}
\label{subsec:mylabel3}
As Brillet-Hall noted, with a single arm interferometer, any variation due 
to anisotropy would have a peak-to-peak magnitude (total change) represented 
by the magnitude of (\ref{eq16}). That is, since an increase in time delay must be 
compensated for by an increase in wavelength (i.e., a decrease in 
frequency),
\begin{equation}
\label{eq17}
\frac{\Delta t}{t}\cong \frac{\Delta \lambda }{\lambda }\cong \frac{-\Delta 
f}{f}\cong \frac{1}{2}\beta ^2\quad \mbox{(total}\;\mbox{change, 
Brillet-Hall)}.
\end{equation}
This contrasts with the Michelson-Morley experiment in which there were two 
perpendicular arms and the peak-to-peak signal would be twice the magnitude 
of (\ref{eq17}). In both experiments, this detected signal would vary with 2X the 
apparatus rotation rate. In the Brillet-Hall experiment the amplitude of the 
fractional change in frequency $\Delta f / f$ of the HeNe laser radiation source 
employed may thus be expected, to lowest order, to be half of the absolute 
value of (\ref{eq17}), i.e
\begin{equation}
\label{eq18}
\frac{\Delta f}{f}\cong \frac{1}{4}\beta ^2\quad \mbox{(amplitude, 
Brillet-Hall)}.
\end{equation}
At the location of the Brillet-Hall test, the earth surface speed is 355 
m/sec. So, for $c =$ 3X10$^{8}$ m/sec one finds $\raise.5ex\hbox{$\scriptstyle 
1$}\kern-.1em/ \kern-.15em\lower.25ex\hbox{$\scriptstyle 2$} \beta ^{2}$ 
= 7.00 X10$^{-13}$ (peak-to-peak). The amplitude of the fractional time 
delay $\Delta t/t$ (= $\vert \Delta f / f \vert $ ) between successive exiting 
waves obtained as the apparatus rotates would be half of this or 3.5 
X10$^{-13}$. This signal should have constant phase angle relative to the 
lab (earth) frame.

\subsection{Billet and Hall: Test Result}
\label{subsec:billet}
The mean value of the anomalous signal Brillet-Hall measured was $\Delta 
f/f\cong 2\times 10^{-13}$ at twice the table rotation rate and constant 
phase angle in the lab. Data varied between --42{\%} and +52{\%} of the 
mean.

The detected mean value is approximately 57{\%} (2/3.5) of that predicted by 
NTO analysis of the signal time delay due to an anisotropy induced by the 
earth surface speed. Considering the experiment covered a range that 
extended to 10$^{-15}$, one could consider this result alone to be quite 
significant. However, there are additional effects on the Brillet-Hall 
signal predicted by NTO analysis that modify the prediction and bring it 
significantly closer to the measured value. These are discussed in the 
following section.

\section{FURTHER NTO EFFECTS ON BRILLET-HALL SIGNAL}
\label{sec:further}
\subsection{Qualitative Considerations}
We emphasize that the effect of the earth surface speed on the speed of 
light in (\ref{eq6}) is due to non-time-orthogonality in a rotating frame and is 
\textit{not} the result of an ``ether wind''. However, it can aid in analysis and 
visualization of the effect described below if one temporarily ignores the 
second order contribution in (\ref{eq6}), and thinks of the effect somewhat 
classically. Then the earth surface velocity \textbf{v} (magnitude $v)$ can be 
effectively considered as a simple vector addition to the usual velocity 
(magnitude $c)$ of a light ray.

In Figure 4 one sees the NTO effect of \textbf{v} in the Brillet-Hall 
experiment when it is transverse to the path of a beam of light. In the left 
side of the figure the portion of the incoming laser beam between the two 
mirrors, which we designate as ``leg 2'', is perpendicular to the direction 
of \textbf{v}. The result is a deflection of the beam by an amount $\Delta 
_{2}$ at the right hand mirror. Since the mirror is aligned at 45$^{o}$ to 
leg 2, the same $\Delta _{2}$ deflection of the beam will occur to the 
beam as it enters, and leaves, the Fabry-Perot interferometer. Hence, the 
peak intensity fringe that would otherwise strike the sensor directly is 
moved to the right in the figure by the amount $\Delta _{2}$.

Similar effects can be seen in the right hand side of the figure for leg 1 
($\Delta _{1}$ deflection) and leg 3 ($\Delta _{3}$ deflection) when the 
apparatus mounting table is turned by 90$^{o}$ relative to the earth 
surface. Note there is an additional effect of a change in angle of 
incidence for the ray as it traverses leg 3. However, as noted in section 
\ref{subsec:brillet}, this has no effect on the angle at which 
constructive interference occurs, and for our purposes, can therefore be 
ignored.

\begin{figure}[htbp]
\centerline{\includegraphics[width=5.59in,height=3.43in]{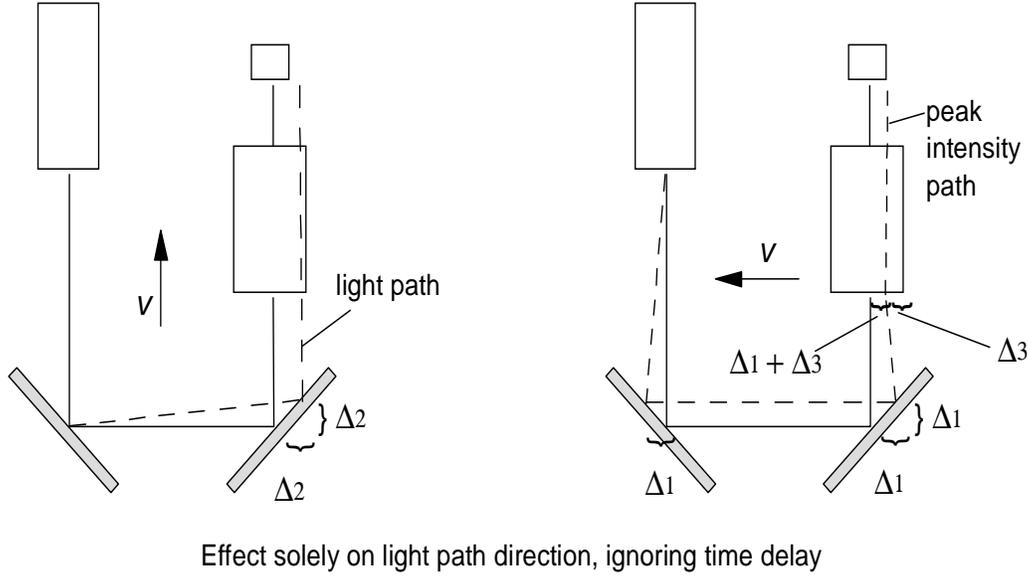}}
\caption{Transverse Effect of Earth Surface Speed on Light Path}
\label{fig4}
\end{figure}

The important point to note is that we now have four separate effects on 
intensity detected by the sensor that are due to the earth surface speed. 
These are

1) Time delay (Section \ref{subsec:mylabel3}),

2) $\Delta _{1}$ from transverse effect on leg 1,

3) $\Delta _{2}$ from transverse effect on leg 2, and

4) $\Delta _{3}$ from transverse effect on leg 3.

\subsection{Quantification of Transverse Effects}
\label{subsec:quantification}
Taking the orientation for \textbf{v} relative to the test table of the left 
side of Figure 4 as $t$ = 0, and using the dimensions for each leg shown in 
Figure 3, the lowest order value for $\Delta _{2}$ is
\begin{equation}
\label{eq19}
\Delta _2 =d_2 \left( {\frac{v}{c}} \right)\cos \omega _T t,
\end{equation}
where $\omega _T $ is the rate at which the table turns in radians/sec.

Similarly, we have for the other two legs,
\begin{equation}
\label{eq20}
\Delta _1 =d_1 \left( {\frac{v}{c}} \right)\cos \left( {\omega _T 
t-\frac{\pi }{2}} \right),
\end{equation}
and
\begin{equation}
\label{eq21}
\Delta _3 =-d_3 \left( {\frac{v}{c}} \right)\cos \left( {\omega _T 
t-\frac{\pi }{2}} \right).
\end{equation}
Note each of these quantities represents the respective displacement of the 
entire fringe pattern at the sensor location to the right of the solid line 
(isotropic peak intensity path) entering the sensor in Figures 3 and 4. 

\subsection{Cumulative Result of NTO Effects}
\label{subsec:cumulative}
From Figure 1, it is apparent that a decrease in wavelength increases the 
peak intensity angle $\alpha $. Further, from Figure 2, one sees that the 
time delay due to anisotropy is similar in effect to a decrease in 
wavelength, and thus also increases the peak intensity angle. The effect of 
time delay on the fringe pattern is shown in the left side of Figure 5, 
i.e., it moves the fringes, including the peak intensity fringe, outward.

\begin{figure}[htbp]
\centerline{\includegraphics[width=5.59in,height=3.43in]{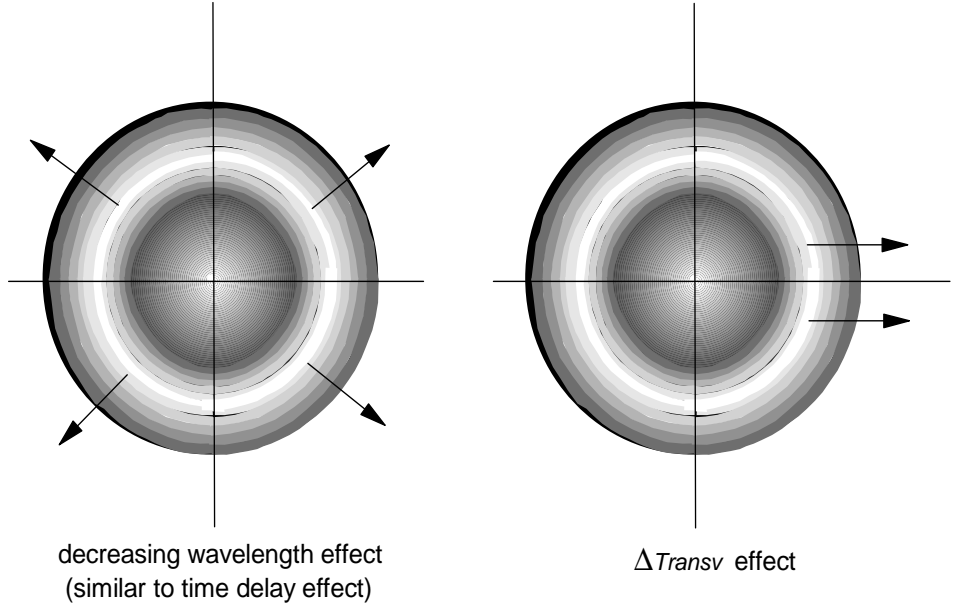}}
\caption{Effects of Interference Fringe Pattern}
\label{fig5}
\end{figure}

The total displacement due to transverse effects is
\begin{equation}
\label{eq22}
\Delta _{Transv} =\Delta _1 +\Delta _2 +\Delta _3 .
\end{equation}
The effect of positive $\Delta _{Transv}$ on the interference fringes is 
depicted in the right side of Figure 5, i.e., it displaces the fringes to 
the right. Negative $\Delta _{Transv}$ displaces the fringes to the left.

The peak intensity angle relation (\ref{eq14}) can be expanded as
\begin{equation}
\label{eq23}
\frac{m\lambda }{2d}=\cos \alpha =\sqrt {1-\sin ^2\alpha } \cong 
1-\textstyle{1 \over 2}\alpha ^2,
\end{equation}
since $\alpha $ is small. Consider, for simplicity, the ideal isotropic case 
in which a local peak intensity is at $\alpha $ = 0 and is directed 
precisely at the center of the sensor. (Imagine the sensor at the origin in 
the left side of Figure 5, light speed is isotropic, and the peak intensity 
is located at the origin.) From (\ref{eq23}) this means
\begin{equation}
\label{eq24}
\lambda _0 =\frac{2d}{m},
\end{equation}
where $\lambda _0 $ is the wavelength in the isotropic case that maximizes 
intensity at the center of the Fabry-Perot interferometer. (For the case 
where $\alpha $ is not initially zero, see Appendix C.) Thus, from (\ref{eq23}) and 
(\ref{eq24}), we have
\begin{equation}
\label{eq25}
\frac{\lambda }{\lambda _0 }=\frac{\lambda _0 +\Delta \lambda }{\lambda _0 
}\cong 1-\textstyle{1 \over 2}\alpha ^2\quad \quad \to \quad \quad 
\frac{\Delta \lambda }{\lambda _0 }\cong -\textstyle{1 \over 2}\alpha ^2.
\end{equation}
Note that $\Delta \lambda $ is always negative, and this makes sense 
physically. A decrease in $\lambda $ means an increase in $\vert \alpha 
\vert $.

Consider (\ref{eq25}) with regard to the time delay only case of Figure 2, i.e., we 
introduce anisotropy into our heretofore isotropic picture, but ignore 
transverse effects. When the Fabry-Perot interferometer is aligned in the 
direction of \textbf{v}, the effect is similar to that of decreased $\lambda 
$. From our analysis of Section \ref{subsec:mylabel3} and (\ref{eq25}), we 
expect the relative change in wavelength to vary with an amplitude of $\beta 
^2/4$ between zero and a negative value of $\beta ^2/2$ at twice the rate of 
test table rotation. From (\ref{eq17}) and (\ref{eq25}) then, the time delay effect on 
wavelength is
\begin{equation}
\label{eq26}
\frac{\Delta t}{t}\cong \frac{\Delta \lambda }{\lambda _0 }\cong 
-\textstyle{1 \over 4}\left( {\frac{v}{c}} \right)^2\left( {\cos \left( 
{2\omega _T t} \right)+1} \right)\cong -\textstyle{1 \over 2}(\alpha 
_{TimeDelay} )^2.
\end{equation}
Trigonometry then gives us
\begin{equation}
\label{eq27}
\left( {\alpha _{TimeDelay} } \right)^2\cong \textstyle{1 \over 2}\left( 
{\frac{v}{c}} \right)^2\left( {\cos \left( {2\omega _T t} \right)+1} 
\right)=\left( {\frac{v}{c}} \right)^2\cos ^2\left( {\omega _T t} \right),
\end{equation}
and the variation in $\alpha _{TimeDelay} $ is
\begin{equation}
\label{eq28}
\alpha _{TimeDelay} \cong \pm \frac{v}{c}\cos \left( {\omega _T t} \right).
\end{equation}
The displacement variation at the sensor location due to time delay alone, 
where $d_{s}$ is the distance from the exit end of the Fabry-Perot etalon to 
the sensor (see Figure 3), is thus
\begin{equation}
\label{eq29}
\Delta _{TimeDelay} \cong (d+d_s )\alpha _{TimeDelay} \cong -(d+d_s 
)\frac{v}{c}\cos \left( {\omega _T t} \right).
\end{equation}
From Figures 2, 4 and 5, we can convince ourselves that $\Delta _{TimeDelay} 
$ at the sensor location due to time delay is 180$^{o}$ out of phase with 
$\Delta _{2}$. Hence, from (\ref{eq19}), we see that the proper sign in (\ref{eq28}) and 
(\ref{eq29}) is the negative one.

The total displacement then is
\begin{equation}
\label{eq30}
\Delta _{Tot} =\Delta _{TimeDelay} +\Delta _{transv} =\Delta _{TimeDelay} 
+\Delta _1 +\Delta _2 +\Delta _3 
\end{equation}
or
\begin{equation}
\label{eq31}
\Delta _{Tot} =-\frac{v}{c}\left( {-(d+d_s )\cos \left( {\omega _T t} 
\right)+d_1 \cos \left( {\omega _T t-\frac{\pi }{2}} \right)+d_2 \cos \left( 
{\omega _T t} \right)-d_3 \cos \left( {\omega _T t-\frac{\pi }{2}} \right)} 
\right).
\end{equation}
This displacement must be corrected for by the servo to maximize intensity. 
That is, the servo must change the frequency, and hence the wavelength, to 
bring $\Delta _{Tot}$ back to zero. In other words, the induced change in 
wavelength must be such that if acting alone (in an isotropic world) it 
would produce an angle $\alpha _{Tot} $ that would move the peak intensity 
fringe by $-\Delta _{Tot} $. This angle is then
\begin{equation}
\label{eq32}
\alpha _{Tot} \cong \frac{-\Delta _{Tot} }{d+d_s }\cong \frac{v}{c}\left( 
{\left( {\frac{d_2 -(d+d_s )}{(d+d_s )}} \right)\cos \left( {\omega _T t} 
\right)+\frac{d_1 -d_3 }{(d+d_s )}\cos \left( {\omega _T t-\frac{\pi }{2}} 
\right)} \right).
\end{equation}
This can be represented as
\begin{equation}
\label{eq33}
\alpha _{Tot} =\frac{v}{c}A\cos \left( {\omega _T t-\phi } \right)
\end{equation}
where
\begin{equation}
\label{eq34}
A=\frac{\sqrt {(d_2 -d-d_s )^2+(d_1 -d_3 )^2} }{d+d_s }
\end{equation}
and $\phi $ is readily determinable. From (\ref{eq25}), and a standard trigonometric 
relation, we have
\begin{equation}
\label{eq35}
\frac{\Delta f}{f_0 }\cong -\frac{\Delta \lambda }{\lambda }=\textstyle{1 
\over 2}\left( {\alpha _{Tot} } \right)^2=\textstyle{1 \over 2}\left( 
{\frac{v}{c}} \right)^2A^2\cos ^2\left( {\omega _T t-\phi } 
\right)=\textstyle{1 \over 4}\left( {\frac{v}{c}} \right)^2A^2\left( {1+\cos 
\left( {2\omega _T t-2\phi } \right)} \right).
\end{equation}
Thus, the fractional change in frequency induced by the servo varies at 
twice the table rotation rate at amplitude
\begin{equation}
\label{eq36}
\left. {\frac{\Delta f}{f_0 }} \right|_{amplitude} =\textstyle{1 \over 
4}\left( {\frac{v}{c}} \right)^2A^2.
\end{equation}
\subsection{Comparison of NTO Theory with Brillet-Hall Experiment}
\label{subsec:comparison}
Brillet-Hall noted that the Fabry-Perot interferometer they used had $d$ = 30.5 
cm, but it is unfortunate that the dimensions $d_{1}$, $d_{2}$, $d_{3}$, and 
$d_{s}$ were not specified. However, from the schematic of their Figure 1, 
and knowing the dimensions of the table illustrated therein, one can make 
reasonable estimates for these values. For example, they appear to have 
values in centimeters which may be reasonably close to ($d_{2}$ appears to be 
between 20 and 25 cm)
\[
d_s =10\quad \quad d_1 -d_3 =25\quad \quad d_2 =22.5.
\]
Then $A^{2}$ = .58 and
\begin{equation}
\label{eq37}
\left. {\frac{\Delta f}{f_0 }} \right|_{amplitude} =.138\left( {\frac{v}{c}} 
\right)^2=2.03\times 10^{-13}.
\end{equation}
As noted in Section \ref{subsec:billet}, the Brillet-Hall anomalous 
signal mean was approximately this value, and like the NTO prediction, had 
both constant phase (value not noted in the article) relative to the lab and 
variation at twice the table rotation rate.

\section{GRAVITATIONAL ORBITS VS TRUE ROTATION}
\label{sec:gravitational}
Why should we get a null signal for the solar and galactic orbital 
velocities, but a non-null signal for the earth surface speed from its own 
rotation? The answer is that bodies in gravitational orbits follow 
geodesics, i.e., they are in "free fall". That is, they are in locally 
inertial (time orthogonal) frames and therefore obey Lorentzian mechanics. 
An object in such a frame feels no force, provided it does not rotate about 
its own axis. There is no experimental means by which one could measure 
(without looking outside at the stars) one's rate of revolution (and hence 
one's speed in orbit).

On the other hand, objects (like the Brillet-Hall or Michelson-Morley 
apparatuses) fixed in "true" rotational frames are held in place by 
non-gravitational forces and do not travel geodesic paths. Via experiment, 
one can actually determine one's speed in such frames in the circumferential 
direction (without looking outside) in an absolute (or Machian) sense.

For example, in true rotation one can measure the motion of a Foucault 
pendulum's arc over time and determine the rate of rotation $\omega $. One 
can also use a mass-spring system to determine the distance $r$ to the center 
of rotation, and thereby calculate the circumferential speed $v=\omega r$ in 
an absolute sense. In gravitational orbit this can not be done. If the 
planet does not rotate about its own axis, there is no movement of a 
pendulum's arc, even thought the planet travels around its orbit. No 
centrifugal acceleration is felt, so one can not use a spring-mass system to 
determine where the center of revolution is. Hence, speed in gravitational 
orbit is indeterminate via experiment.

Thus, the earth surface speed is that of a true rotational system; orbital 
speed is not. This has at least two immediate ramifications. First, if for 
true rotation one can measure circumferential speed via one experiment 
(pendulum, spring, mass), then why not by another (Michelson-Morley or 
Brillet-Hall)? Second, the non-time-orthogonal factor (see off diagonal term 
in (\ref{eq3})) is a direct function of $\omega $ and $r$. Measuring those, we 
effectively measure the degree of non-time-orthogonality, and hence the 
degree of anisotropy. 

Hence, the usual form of relativity should hold for gravitational orbits and 
we should expect null Michelson-Morley and Brillet-Hall results for orbital 
speeds, which is just what we measure. But the NTO form of relativity theory 
should be applied for the earth surface speed due to rotation about the 
earth's axis. This issue is treated in detail in 
Klauber\cite{Klauber:4}.

Note that the anisotropy in light speed is predicted to exist on the earth 
provided the earth is rotating, though it would be undetectable in practice 
unless the experimental apparatus turned with respect to the earth's 
surface, thereby enabling speed comparisons in different directions (e.g., 
E-W then N-S) within the rotating earth frame itself. Although other 
tests\cite{Braxmaier:2002} have been performed with accuracy comparable to, 
or better than, that of Brillet-Hall, none of these turned the test 
apparatus relative to the earth surface.

\section{SUMMARY AND CONCLUSIONS}
\label{sec:summary}
Non-time-orthogonal analysis of relativistically rotating frames predicts a 
signal that is in accord with the anomalous non-null signal on the order of 
10$^{-13}$ found in the Brillet-Hall experiment. Contributions to the signal 
come from time delay and three separate transverse effects on wave 
propagation. Both the predicted signal and the measured signal have fixed 
phase in the lab frame and vary with twice the test apparatus rotation rate. 
Signal magnitude and phase depend on particular dimensions between 
components of the test apparatus, which are unknown, though estimable. 
Reasonable estimates for these dimensions result in a prediction that is 
well within the range of data found via experiment.

NTO analysis predicts a signal that is dependent on the surface velocity of 
the earth due to rotation about its axis, but none due to the velocity of 
the earth in gravitational orbit. This is in agreement with all relevant 
experiments. This result is due to off diagonal time-space terms that arise 
in the metric for true rotating frames (where angular velocity and 
centrifugal acceleration can be measured). No such off diagonal terms exist 
in the metric for bodies in the free fall frames of gravitational orbit.

This result contrasts with the null prediction of the usual relativistic 
analysis approach, utilizing local co-moving Lorentz frames. The 
Brillet-Hall measurement, GPS data, and the analysis herein raise 
significant questions relating to relativity theory and the nature of the 
spacetime continuum. (Can time truly be non-orthogonal to space, and if so, 
what ramifications are there for physical measurements?) The author hopes 
that someone will soon repeat the Brillet-Hall experiment (with apparatus 
rotating with respect to the earth surface) and provide definitive answers 
to these questions.

\section*{APPENDIX A: THE FABRY-PEROT INTERFEROMETER}
For reference, this appendix provides a derivation of the well known 
Fabry-Perot constructive interference condition for thin films\footnote{ 
This article does not delve into certain intricacies of geometrical optics 
in rotating frames, particularly with regard to reflection and 
half-reflection, and in this sense, derivations herein are approximate.}.

Figure 1 depicts the Fabry-Perot interferometer with two partially 
transmitting glass (silica) mirror sections mounted a distance $d$ (the 
``etalon'' length) apart. A light ray enters from the left at angle $\alpha 
$ to the horizontal, is refracted through the first glass section, and then 
is partially reflected at the surface of the second glass section. The 
reflected ray is reflected once again at A, then partially refracted at B, 
and upon exiting the right side of the second glass section interferes with 
the original refracted ray. Two phase changes of 180$^{o}$ occur (at the 
first reflection and the second reflection, point B) so the net phase change 
due to reflection upon exiting the right hand glass section is zero. The 
reflected portion of the ray at B leads to a third ray, as well as 
subsequent rays, exiting the RHS. Each successive such ray is diminished in 
intensity from the prior one, and the interference behavior of all of these 
rays together parallels that of the first two shown in Figure 1.

Note that for constructive interference, we must have
\begin{equation}
\label{eq38}
2AB-CD=m\lambda ,
\end{equation}
where $m$ is an integer. The following trigonometry is straightforward.
\begin{equation}
\label{eq39}
AB=\frac{d}{\cos \alpha }
\end{equation}
\begin{equation}
\label{eq40}
l=d\tan \alpha 
\end{equation}
\begin{equation}
\label{eq41}
CD=2l\sin \alpha =2d\tan \alpha \sin \alpha 
\end{equation}
Using the above values for AB and CD in (\ref{eq38}), one finds
\begin{equation}
\label{eq42}
\frac{2d}{\cos \alpha }-2d\tan \alpha \sin \alpha =m\lambda .
\end{equation}
Multiplication by cos$\alpha $ yields
\begin{equation}
\label{eq43}
1-\sin ^2\alpha =\cos \alpha \frac{m\lambda }{2d},
\end{equation}
or
\begin{equation}
\label{eq44}
\frac{m\lambda }{2d}=\cos \alpha .
\end{equation}
\section*{APPENDIX B: GALILEAN ROUND TRIP TIMES IN TRANSLATING FRAMES}
\label{sec:appendix}
The general relation for determining measured round trip travel times for 
different speeds in each direction is
\begin{equation}
\label{eq45}
t_{RT} =\frac{l}{v_+ }+\frac{l}{v_- }
\end{equation}
where $l$ is the distance traveled in one direction measured by standard meter 
sticks, $v_{+}$ is the physical (measured) speed in the outward direction and 
$v_{-}$ is the physical return speed.

Using (\ref{eq45}), the pre-relativistic ether approach to light travel times in 
directions parallel and perpendicular to velocity of translation through the 
ether\cite{See:1965} yielded, to second order,
\begin{equation}
\label{eq46}
t_{RT,perpend} \cong \frac{2l}{c}\left( {1+\textstyle{1 \over 
2}\frac{v^2}{c^2}} \right)
\end{equation}
and
\begin{equation}
\label{eq47}
t_{RT,parallel} \cong \frac{2l}{c}\left( {1+\frac{v^2}{c^2}} \right).
\end{equation}
The difference between these is
\begin{equation}
\label{eq48}
\Delta t_{\scriptsize \begin{array}{l}
 RT,Galilean \\ 
 moving\,frame \\ 
 \end{array}} \cong \frac{2l}{c}\left( {\textstyle{1 \over 
2}\frac{v^2}{c^2}} \right)=(t_{RT,no\;motion} )\left( {\textstyle{1 \over 
2}\frac{v^2}{c^2}} \right)=\textstyle{1 \over 2}\beta ^2(t_{RT,no\;motion} 
)
\end{equation}
The fractional difference in round trip light travel time between the two 
paths equals $\raise.5ex\hbox{$\scriptstyle 1$}\kern-.1em/ 
\kern-.15em\lower.25ex\hbox{$\scriptstyle 2$} \beta ^{2}$.

This value is reflected directly in the predicted (for classical ether) 
change in interference fringing in the Michelson-Morley experiment and in 
the change in frequency of the Fabry-Perot interferometer in the 
Brillet-Hall experiment. This, of course, presumes that motion in each case 
can be modeled as translational and Galilean.

It is significant that the Michelson-Morley experiment was not sensitive 
enough to detect any effect from the earth surface speed, though the 
Brillet-Hall experiment was.

\section*{APPENDIX C: CASE OF INITIAL PEAK INTENSITY AT $\alpha \ne 0$}
\label{sec:mylabel2}
Consider the case where the sensor is not located at the precise centerline 
of the Fabry-Perot interferometer, such that the local peak in intensity 
striking the sensor is not at $\alpha $ = 0. That is, $\alpha _0 \ne 0$, 
though still small. Then
\begin{equation}
\label{eq49}
\frac{m\lambda _0 }{2d}=\cos \alpha _0 \cong 1-\textstyle{1 \over 2}\alpha 
_0 ^2
\end{equation}
yields the wavelength $\lambda _{o}$ for initial peak intensity at the 
sensor. As the apparatus turns, due to anisotropy, the servo changes the 
wavelength to maintain constant $\alpha _{o}$. The extremum value for this 
wavelength is $\lambda _{1}$, and this corresponds to what the peak 
intensity angle $\alpha _{1}$ would be for such a wavelength if the 
apparatus had not turned. Thus,
\begin{equation}
\label{eq50}
\frac{m\lambda _1 }{2d}=\cos \alpha _1 \cong 1-\textstyle{1 \over 2}\alpha 
_1 ^2.
\end{equation}
From (\ref{eq49}) and (\ref{eq50}),
\begin{equation}
\label{eq51}
\frac{\Delta \lambda }{\lambda _0 }=\frac{\lambda _1 -\lambda _0 }{\lambda 
_0 }=\frac{\lambda _1 }{\lambda _0 }-1=\frac{\cos \alpha _1 }{\cos \alpha _0 
}-1\cong \frac{1-\textstyle{1 \over 2}\alpha _1 ^2}{1-\textstyle{1 \over 
2}\alpha _0 ^2}-1\cong -\textstyle{1 \over 2}\left( {\alpha _1 ^2-\alpha _0 
^2} \right).
\end{equation}
For 
\begin{equation}
\label{eq52}
\alpha _1 =\alpha _0 +\Delta \alpha ,
\end{equation}
(\ref{eq51}) becomes
\begin{equation}
\label{eq53}
\frac{\Delta \lambda }{\lambda _0 }\cong -\textstyle{1 \over 2}\left( 
{(\Delta \alpha )^2+2(\Delta \alpha )\alpha _0 } \right).
\end{equation}
The first term on the right side will cause a second order variation in 
$\Delta \lambda $ at twice the table rotation rate. (See (\ref{eq26}) to (\ref{eq28}).) 
The last term should cause a variation at the table rotation rate with a 
magnitude dependent on the initial condition $\alpha _{o}$.

In this context, it may be significant that Brillet-Hall detected another 
non-null signal of magnitude $\sim $1.2X10$^{-12}$ at the table rotation 
rate. They attributed this signal to gravitational stretching of their 
apparatus, but in light of (\ref{eq53}), one might wonder if the cause lies 
elsewhere. A repetition of their test would tell us a lot.

\end{document}